\documentclass[letterpaper]{jpconf}
\usepackage{graphicx}
\usepackage{amsfonts,amssymb}
\newcommand{\eq}[1]{Eq.~(\ref{eq:#1})}
\newcommand{\be}{\begin{equation}}
\newcommand{\ee}{\end{equation}}
\newcommand{\ba}{\begin{array}}
\newcommand{\ea}{\end{array}}
\newcommand{\bea}{\begin{eqnarray}}
\newcommand{\eea}{\end{eqnarray}}

\newtheorem{conjecture}{Conjecture}
\newenvironment{conjecturetwop}{
\smallskip

\noindent {\bf Conjecture 2}$^\prime$
\itshape}{\normalfont\smallskip}

\def\bh{{\bar h}}
\def\bt{{\bar t}}
\def\bL{{\bar L}}
\def\vev#1{\langle{#1}\rangle}

\def\BZ{\mathbb{Z}}
\def\BR{\mathbb{R}}

\def\CL {{\cal L}}

\def\CH {{\cal H}}

\begin{document}
\title{Spaces of Quantum Field Theories}

\author{M R Douglas}

\address{Simons Center for Geometry and Physics\\
Stony Brook University, Stony Brook NY 11794}
\address{IHES, Le Bois-Marie, Bures-sur-Yvette France 91440}

\ead{douglas@max2.physics.sunysb.edu}

\begin{abstract}
The concept of a ``space of quantum field theories'' or ``theory space'' 
was set out in the 1970's in work of Wilson, Friedan and others.
This structure should play an important role in organizing and classifying QFTs, and in the
study of the string landscape, allowing us to say when two theories are connected by
finite variations of the couplings or by RG flows, when 
a sequence of QFTs converges to another QFT,
and bounding the amount of information needed to uniquely specify a QFT, enabling us
to estimate their number.  As yet we do not have any definition of theory space which
can be used to make such arguments.

In this talk, we will describe various concepts and tools which should be developed
for this purpose, inspired by the analogous mathematical problem of studying the
space of Riemannian manifolds.  We state two general conjectures about the 
space of two-dimensional conformal field theories, and we 
define a distance function on this space, 
which gives a distance between any pair of theories, whether or not they
are connected by varying moduli. 

Based on talks given at QTS6 (University of Kentucky), Erice, Texas A\& M, and Northwestern University.
To appear in the proceedings of QTS6.

\end{abstract}

\section{Introduction}

Quantum field theory (QFT) is a remarkably successful physical
framework, describing particle physics, critical phenomena, and
certain many-body systems.  Superstring theory is also closely based
on QFT.  A central problem of theoretical physics is to classify the QFTs and understand
the relations between them.

After more than fifty years of effort, we have no complete classification, nor is such a thing
even on the horizon, except for a few special subclasses of theories.  In the broadest terms,
we can classify QFT's by the dimension of space-time, and the presence of continuous symmetry
and/or supersymmetry.  Within such a class, to get a well posed problem, we should place some
upper bound on the total number of degrees of freedom, under some precise definition.

The most basic question of classification would then be: is there any sense in which the total set
of QFTs in this class is finite, or countable, or compact, or something we can in any way put
definite limits on?  To put this a different way, can we make a list, say of Lagrangians for
definiteness, in which every QFT is guaranteed a place on the list?  In itself this need not be
the final answer -- an entry on the list might need to satisfy further
consistency conditions to be a QFT, or several entries might describe the same QFT.
Still, given such a starting point, we could continue by addressing these difficulties
to systematically map out the space of QFTs.
But, except for a few special classes, we have no such list, and no answer to this question.

Now, given
sufficiently restrictive assumptions, it can happen that some finite subset of choices
of field content and coupling constants fixes the entire theory, making the classification
problem tractable.
For example, if we are interested
in four-dimensional theories with $16$ supercharges,\footnote{
In this talk, we only discuss unitary field theories without gravity.}
there are good arguments that the only examples are the supersymmetric Yang-Mills theories.
These are determined by a choice of gauge group $G$ (a compact Lie group) and a
complexified gauge coupling for each group factor, with certain known equivalences
between theories (S-dualities).  

Another class of QFTs which has been classified is the ``minimal CFTs,'' the
two-dimensional conformal field theories with central
charge $c<1$.  Their classification is a bit more complicated, with an ``algebraic'' part (the
representation theory of the Virasoro algebra) determining the spectrum of primary
fields, and a ``global'' part which determines the list of closed operator algebras satisfying
modular invariance (see \cite{Gannon:1999bi,Cappelli:2009xj} for an overview
of this classification).

While there are a few more results of this type, the general classification problem, even for
cases of central interest such as $d=4$, $N=1$ supersymmetric theories, seems so distant that
it is not much discussed.  This is because we have no answer to the question we posed above.
More specifically, because we do not sufficiently well understand
the relation between the description of a QFT (such as a bare Lagrangian)
and the actual observables,
it is not known how to reduce the problem  to finite terms.  
 
To illustrate this point, let us consider what might appear to be a simple analog to the 
minimal CFTs, namely the 2d CFT's with $c<2$.  As is well known, the  
representation theory of the Virasoro algebra does not get us very far here; the number of
primary fields grows very quickly with dimension (superpolynomially) and we do not have
any way to reduce this large amount of data to a manageable subset.  While one can consider
larger algebras and rational CFTs, this amounts to looking under a rather
well inspected lamppost; we know that
there are nonrational CFTs, and that crucial aspects of the problem (such as the fact that
CFTs can have moduli) are not respected by this simplification.

There is no reason why we must take an algebraic approach, and
one might try to describe the space of $c<2$ CFTs in other ways.
For example, one could start by considering the Lagrangians with two
scalar fields and a nontrivial potential.  By the $c$-theorem, we know all such theories will flow to
CFTs (often trivial) with $c< 2$.  While these could be listed, there is no reason to think this includes
all the $c<2$ theories.  An example of a
$c<2$ theory which probably cannot
be obtained this way is the direct sum of an Ising model ($c=1/2$) with two tricritical
Ising models ($c=7/10$).  

This particular theory can be obtained using three scalar fields, and 
it is a reasonable hypothesis that any $c<2$ theory could be obtained in a similar way,
as a flow from a starting point with more scalar fields. 
If this were the case, we could imagine solving the classification problem by simply listing Lagrangians,
but we would need one more ingredient for this list to have an end:
an ``inverse $c$-theorem'' that tells us that all theories with $c< 2$ 
can be obtained as RG flows from a set of theories with at most $N$ scalar fields.  
At present we do not know whether this is so, nor is there any conjecture for the $N$ 
required for such a claim to hold.

This is an example of a fairly concrete approach to classification, which might or might not work.
But one of the points I want to make in this lecture is that  
we (physicists) have too simple an idea of what a classification should be.  
I believe we need to broaden this idea to make progress, in ways analogous to
those which were developed to
address similar problems in mathematics.

Intuitively, one classifies
a set of mathematical objects by listing them, and giving an algorithm which, given an object
$X$, determines which one it is.  The compact Lie groups are a paradigmatic example in which
the objects are classified by discrete structures.  As with the minimal CFTs, this classification has an
algebraic part, the classification of Lie algebras (say in terms of Dynkin diagrams), and a global
part, the classification of discrete subgroups which can appear as the fundamental
group.  In other
cases, one has families of objects depending on parameters.  One then needs to understand 
the moduli spaces in which these parameters take values; 
familiar examples are the moduli spaces of Riemann surfaces.  

These are cases in which the classification program was successful.  On the other hand, many
other superficially analogous mathematical problems have not been solved in such concrete terms, and we will
argue that this includes problems quite analogous to classification of QFT.   This did not stop the
mathematicians, but rather inspired the development of other ways to think about classification problems,
more abstract than listing objects or defining moduli spaces.  Some are familiar in physics, such as the
idea of a topological invariant.  Others are not, including most of the ideas of metric geometry.
\cite{Gromov}

Before proceeding, let us recall the best existing ``classification'' of QFTs, which is based on perturbation
theory.  A large set of QFTs (perhaps all of them)
can be defined by quantizing a classical action, say by using a regularized functional integral,
computing in perturbation theory,
and performing renormalization.  For these theories, the classification problem has three parts.
First, we need to enumerate actions and the couplings they depend on.  We also need to
identify equivalences between theories and redundant couplings.
We then need
to study the renormalization group and decide which operators are relevant and marginal,
so that we can take the continuum limit.   
In perturbation theory, these steps
are by now fairly well understood.

Finally, we need to understand the relation between the perturbation theory and the actual
observables of the theory.  In some cases, this is fairly clear -- quantities computed in
perturbation theory are asymptotic to the true correlation functions or S-matrix elements
as the couplings go to zero.  If the physics at finite coupling is qualitatively similar to the
limiting free theory, one can essentially regard the coefficients of the perturbative expansion as 
physical observables.  If these agree between two theories, they are the same; if they are
``close'' in some sense the theories are close, and so on.  

Although this map from the action to the observables is somewhat complicated, as physicists
we learn how to work with it, and this is our working definition of a ``space of QFTs.''  Thus the
stable particles correspond to fields, certain leading coefficients in scattering amplitudes or 
correlation functions correspond to couplings, and so on.   The resulting moduli space of
QFTs is a space with one component for each choice of field content, and with each 
component parameterized by a finite set of
couplings in the Lagrangian.

One problem with this working definition is that it is based on
perturbation theory, and there are many QFTs for which this is
not a good description.  Formally, we might
distrust perturbation theory because the bare
coupling is large, or because a coupling is relevant and becomes large at low energy.
Now in itself this does not prove that the physics is qualitatively different from the
picture given by perturbation theory; an example where it is not is
the 2d Landau-Ginzburg models ({\it i.e.}, scalar fields with a potential with isolated critical points).

But in many other cases we know that it is.  The perturbative
excitations might be confined, so that only composite objects are observable.  Even when the
perturbative excitations are not confined, there might be solitons which are lighter, and in a physical
sense are more fundamental.  
These thoughts lead to the concept of duality equivalences, and the idea that a large part of
theory space might be described by patching together perturbative descriptions which are valid
in different regimes.\footnote{
Perhaps the first significant use of this idea was in \cite{Seiberg:1994rs}.
It was a common thread throughout the mid-90's, as illustrated by \cite{Vafa:1997pm,Douglas:1996vj}.}

Of course, there might be other regions where no perturbative description
is valid.  Indeed, we do not know whether we yet know about all the types of QFT, or whether we 
have all of the concepts we will need to 
describe them.  One test is that a local description of theory space must be complete,
in the sense that any relevant or marginal perturbation keeps us within the space.  One must also
ask whether there are disconnected components of theory space, not obtained by perturbing the
theories we know about.  To address this question, one's basic definitions probably must place the
QFTs within some larger set with a simpler description.

These are attractive questions and ideas, which are part of the general study of the
space of QFTs.  This concept seems to have come out of statistical field
theory; it was discussed in Wilson and Kogut's famous review \cite{Wilson:1973jj}
and many works since.  At least as an intuition, it provides a framework for a good deal of
work on QFT.  Nevertheless, we do not know very much about it, as illustrated by the question
of our introduction, or more precise questions such as these:
\begin{itemize}
\item Is theory space connected (by paths)?  Given two QFTs $X$ and $Y$, are they in the
same component?  If so, is the path between them ``finite length'' or ``infinite length'' ?
\item Consider a sequence $X_i$ of CFTs; what kind of limits can it have?  
\cite{Kontsevich,Roggenkamp:2003qp,Roggenkamp:2008jm,Soibelman}
One knows that the large volume limit is at infinite distance, is this the only infinite 
distance limit?  Some finite distance limits, such as the conifold point, violate standard
axioms of CFT (correlation functions can diverge).  Can we be more precise about what is
happening and classify these limits?
\item How ``big'' is theory space?  Can we show that the number of QFTs satisfying some
conditions is finite, and estimate the number?
\item How many parameters (or measurements) do we need to characterize a QFT?  At 
weak coupling, we can answer this, but more generally?
\end{itemize}

In trying to make these ideas and questions precise, one runs up against unsolved problems.
Certain necessary concepts and tools have not yet been developed, and another goal
of the talk is to explain what these are.

In previous talks on this subject \cite{SimonsYang}
I expressed this point with the phrase that ``to study the space
of Xs'' (here meaning QFT), one must have a definition of an X.  While I would still maintain this,
having the definition is not enough; there are other equally important aspects of the problem.
Let me list a few, which I will explain below:
\begin{itemize}
\item Topology and metric on theory space
\item Embedding theorems
\item Weak QFT
\end{itemize}
In explaining these ideas, we will follow the basic
analogy between the classification of QFT and the
classification of Riemannian manifolds, {\it i.e.} manifolds with metrics,
which first emerged in Friedan's work on renormalization of 2d sigma models 
\cite{Friedan:1980jm}.

\section{Topology and metric on theory space}

The next point I want to make in this talk is that it is useful to make precise definitions of 
when two QFTs are ``close.''  To give the basic idea, suppose we regard a QFT as defined by a
list of all of the observables -- S-matrix
elements, or operator dimensions and o.p.e. coefficients.  We then say that two theories are
at a distance $\epsilon$ in ``theory space'' if the largest difference between these coefficients
is $\epsilon$.  Of course, we could use this to define a distance in terms of the other definitions
of QFT.  

Now, one of the basic problems with regarding a QFT as defined by the list of all of its observables,
is that since there are
an infinite number of operators with an infinite number of correlation functions,
there are an infinite set of observables, and this picture of theory space may seem rather unconstrained
and useless.  Of course, a first step in the discussion is to reduce this to simpler terms, and for 2d CFT
it is known that we can derive all the other correlation functions from the three-point functions of
primary fields.  However, this is still an infinite amount of data.

This should not stop us.  On reflection, one already has a milder version of the same
sort of problem when thinking about spaces
of ordinary functions, even for
a real analytic function of a single real variable.  Such a function is characterized
by its Taylor expansion around a given point, an infinite set of coefficients.  And, indeed, there is no
{\it a priori} upper bound on the number of coefficients one needs to describe such a function.
Nevertheless this point does not stop us; we have a good intuition for how to work with functions,
and in ``normal physics problems'' we can find finite dimensional spaces of functions which are good
enough approximations to describe the functions of interest.

Our goal is to develop the analogous intuition here, which tells us what finite dimensional
subset of this data is needed to actually determine a particular QFT.  Again, there is a simple
conceptual answer to this question -- if we are perturbing around a given theory, we need to know the
couplings for the relevant and marginal operators, a finite subset.  The problem is to say this
in some way which, unlike the standard discussions, does not depend so drastically on 
which theory we are perturbing around.
To do this, we need to formulate a few basic concepts, such as what it means to ``perturb around
a given theory,'' in a more general way.

Let us be a bit more concrete.  Suppose we have a set of QFTs $T(\alpha)$, depending on
various discrete or continuous parameters $\alpha$.
Each QFT has a space of local operators $\phi_i(x)$, and we grant that we
know the $n$-point correlation functions for all the theories.  The problem is to
define a distance between pairs of theories, denoted
\be
d(T_1,T_2) ,
\ee
which satisfies the axioms of a distance (or metric space\footnote{
In parts of the mathematics literature, such a distance function is referred to as a metric;
the metric tensor familiar from general relativity is referred to as a  Riemannian metric.}): 
\bea
&d(T_1,T_2) \ge 0 \qquad &(\mbox{\rm with equality only if}\, T_1\cong T_2).\\
&d(T_1,T_2) = d(T_2,T_1), &\label{eq:symmetry} \\
&d(T_1,T_2) + d(T_2,T_3) \ge d(T_1,T_3) \qquad &\mbox{\rm (the triangle inequality).} \label{eq:triangle}
\eea

One can keep some concrete sets of QFTs in mind which to test a definition.  The first is the
two-dimensional sigma models with target space a torus of dimension $d$, a flat
metric $g_{ij}$, and constant $B$ field.  
These QFTs are CFTs and their moduli space is well known to be %\cite{Narain}
\be
SO(d,d;\BZ) \backslash SO(d,d;\BR) / SO(d;\BR) \times SO(d;\BR)
\ee

The second is the $c=1$ CFTs, which were classified in \cite{Ginsparg:1987eb}.  The moduli
space of such theories consists of several disconnected components.
One is made up of two half-lines parameterizing the radius
of the target spaces $S^1$ and $S^1/\BZ_2$, with the endpoint of the second half-line identified
with a particular point on the first half-line.  There are also three isolated 
``exceptional'' theories
which can be defined as quotients
of the $SU(2)_1$ WZW model by the $E$-type discrete subgroups of $SU(2)$.

\subsection{Zamolodchikov distance}

There is a standard Riemannian metric on any moduli space on CFTs, the Zamolodchikov metric 
\cite{Zamolodchikov:1986gt,Kutasov:1988xb}.
It is defined as follows: a tangent vector $t$ to the space of theories at a theory $T$ corresponds to
a marginal local operator $O_t$ in the theory; the length of a tangent vector $t$ is then
\be
||t||^2_{Zam} \equiv \vev{O_t(0) O_{t}(1)}_T .
\ee
This corresponds to the metric on field space in a space-time effective action for
string theory compactified on $T$.

A Riemannian metric can be used to define a standard distance function -- the distance
between $T_1$ and $T_2$ is the length of the shortest path connecting them. 
In the special case of the torus target spaces,
the Zamolodchikov metric is homogeneous, so any pair of theories $T_1$ and $T_2$
is connected by a unique shortest geodesic which can easily be found explicitly.
We then take $d(T_1,T_2)$ to be the length of this geodesic; familiar geometric arguments
show that this satisfies the stated axioms.

So far all this will probably be familiar, and the reader may be expecting  us to propose 
the straightforward generalization of this.  Any moduli space of CFTs will have a 
Zamolodchikov metric, and we can take the distance $d(T_1,T_2)$ to be the length of 
the shortest path between them.  Again, this will satisfy the axioms.

While this metric is certainly very important in CFT and string theory, 
if we consider it as a candidate distance to use in defining theory
space, it is not really what we want.  The main problem with it is simply that it does not define a distance
between every pair of CFTs (much less QFTs).  For example, the $c<1$ minimal models have
no moduli and are thus isolated points in theory space.  Consider the case $c=4/5$; there
are two unitary partition functions, one diagonal and the other that of the three-state Potts model.
These are not connected within the space of CFTs, so they have no distance in this sense.  But we do
not want to say they are at infinite distance, as they can both be obtained as
limits of spin systems related by simply varying bare couplings.  
Rather, a definition of the distance between them
must be based on other concepts.  

Disconnected components of the space of CFTs, such as the exceptional $c=1$ theories, are common.
A good definition of
theory space must be able to handle them.

\subsection{Distance between CFTs}

Let us propose a different definition of distance which can be applied to any pair of two-dimensional
CFTs with the same central charge $c$.  As explained in the founding paper \cite{Belavin:1984vu}, using the
state-operator correspondence and the operator product expansion, all $n$-point functions can
be expressed in terms of the spectrum of operator dimensions $(h_i,\bar h_i)$,
and the $3$-point functions
\be
C_{ijk} = \vev{\phi_i(z_1) \phi_j(z_2) \phi_k(z_3)}
\ee
for some fixed $z_1,z_2,z_3$,
or equivalently the o.p.e. coefficients.\footnote{
To simplify the discussion here,
we normalize all the operators to have unit two-point function.  This is actually
not the best choice when considering a general family of theories, as it can lead to spurious 
divergences in the o.p.e. coefficients, but we will discuss this point elsewhere.}
Thus, the data we need to describe a CFT is
this (admittedly infinite) list of numbers $(h_i,\bar h_i,C_{ijk})$.  If we can define a distance purely in terms of
this data, it will apply to any pair of CFTs.\footnote{
Since there are isospectral CFTs, for example $E_8\times E_8$ and $Spin(32)/\BZ_2$ at level one,
the distance must depend on $C_{ijk}$.}

Let us regard this data for a theory $T$ as defining a vector $V(T)$ in the space of infinite sequences
of real numbers; then we
could define the distance between a pair of theories $T_1$ and $T_2$ as the distance between the
corresponding vectors.
Furthermore, since the space of sequences of real numbers is a linear space, we can define this
distance in terms of a norm, so
\be \label{eq:normdist}
d(T_1,T_2) = ||V(T_1)-V(T_2)|| .
\ee

A standard class of such norms are the $\ell_p$ norms,
\be
||V||_p = \left( \sum_i |v_i|^p \right)^{1/p} ,
\ee
where $p\ge 1$ is a real number.  One can also take the limit $p\rightarrow\infty$ to get
\be
||V||_\infty = \max_i |v_i| .
\ee
Thus, by choosing some $p$, we get a distance between $T_1$ and $T_2$.

We will also need to choose a normalization convention for the vectors $V$.  One natural 
choice would be to take $||V||=1$, but with this choice the maximum distance between
any pair of theories will be $2$, even in cases such as the large radius limit which we would
have thought should run off to infinite distance.
Later we will propose a different choice.

This style of definition has many disadvantages compared to the Zamolodchikov metric.
It is less natural -- we have to choose $p$, and we are combining
the various CFT data in a somewhat arbitrary way.  At first sight, it does not 
correspond to any observable in string theory.
On the other hand, it has the great advantage that it defines a distance between
{\bf every} pair of CFTs, since every
CFT corresponds to a vector $V(T)$.

\subsection{Symmetries and dualities}

Before discussing the appropriate choice of $p$, we need to deal with
some problems with the definition as stated.  One evident problem is that the vector $V(T)$
is not uniquely defined by the CFT, since it depends on a choice of basis.  We might
try to state a prescription for ordering the basis, such as taking the lowest dimension operators
first.  However, if there are operator degeneracies, and especially if the CFT has symmetries,
such prescriptions will not lead to a unique choice of $V$.

A better way to deal with this problem is as follows.
We consider the Hilbert spaces for the two theories $\CH_1$ and $\CH_2$, and
the set of 
unitary transformations $U:\CH_1\rightarrow \CH_2$.  Any such unitary
transformation determines a corresponding linear transformation on the Hamiltonian
$L_0+\bar L_0$, the o.p.e. coefficients $C_{ijk}$, and on their matrix elements in an
explicit basis.  Thus, it defines an action on the space of linear sequences $V(T)$.  Using
this, we take the previous definition and minimize over all possible
choices of $U$,\footnote{
Note that $U$ is not a general unitary matrix acting on the space of vectors $V$ (in which case the
minimum would be determined by the lengths $||V(T_1)||$ and $||V(T_2)||$), 
since it is constructed as a tensor product of unitaries acting on $\CH$.
Also, one should allow partial isometries $U_1,U_2$ embedding $\CH_1$, $\CH_2$
into a larger Hilbert space.}
\be \label{eq:Udistance}
d(T_1,T_2) = \min_U ||U(V(T_1))-V(T_2)|| .
\ee
While this obviously satisfies the other axioms, we need to check
the triangle inequality.  This can be seen by writing \eq{triangle} as
\be \label{eq:triangle-proof}
 ||U_{12}(V(T_1))-V(T_2)|| 
+ ||U_{32}(V(T_3))-V(T_2)|| \ge \min_{U'} ||U'\cdot U_{12}(V(T_1))-U_{32}(V(T_3))|| .
\ee
where $U_{12}$ and $U_{32}$ are the choices minimizing \eq{Udistance}.  Since the $U$'s all come 
from unitaries acting on $\CH$, minimizing over $U'$ in this expression is the same as minimizing
over $U$ in the expression \eq{Udistance} for $d(T_1,T_3)$.  Now, already for $U'=1$, this
inequality is true, since it is just the ordinary triangle inequality.  But the additional step of
minimizing over $U'$ can only decrease the right hand side, respecting the inequality.

In practice, if $T_1$ and $T_2$ are not very close, it may be difficult to find the minimizing $U$,
and in this sense the definition is nonconstructive.  Nevertheless it is a standard mathematical
definition (used to define the Gromov-Hausdorff distance between manifolds) and
it has a number of advantages.  Most importantly,
it takes account of symmetry and duality relations between theories.  Suppose the theories $T_1$
and $T_2$ are T-dual; then there will exist a unitary transformation $U$ such that the distance
$d(T_1,T_2)=0$, as it should be.

\subsection{Problems with high dimension operators}

While this definition is better than the first one, it still does not work.  Consider the example of a
free boson with radius $R$; one set of operators in the two theories are the winding
operators
\be
V_n = :\exp in(\phi_L-\phi_R): \, , \qquad n\in\BZ
\ee
with dimension $h=n^2 R^2$.  Suppose we consider a pair of theories related
by a small variation of $R$.  Clearly
the operator $V_n$ in the first theory will correspond to $V_n$ in the second, so we will have
\be
h^{(R_1)} - h^{(R_2)} = n^2 \left( R_1^2 - R_2^2 \right) .
\ee
Thus, for any nonzero $R_1-R_2$, as $n$ becomes large this difference
will become arbitrarily large, and the distance defined this way (for any $p$) will be infinite.

Looking at other examples, this problem is very general.  For any definition of this type to
make sense, 
the effect of varying high dimension operators and their o.p.e.'s must be suppressed.

Conversely, if there were to exist pairs of CFTs which agreed in all the o.p.e.'s of low
dimension operators, and only differed for operators with large $h$, we would not
be able to define a space of CFTs.  Intuitively, we do not expect this, because the structure
of the theory is determined by the relevant and marginal operators, but we do not know of
any precise claims to this effect.  

One step towards making this precise would be to demonstrate
the following
\begin{conjecture} \label{q:one}
There exists an $H$, growing linearly with $c$, such that if two CFT's 
agree in their spectrum and three-point functions for
all operators with $h\le H$ and $\bar h\le H$, they must be the same CFT.
\end{conjecture}
Although the intuition based on relevant and marginal operators suggests $H=d$,
this need not be the case.  A related question which comes up in gauge-gravity duality \cite{Witten:2007kt},
is to find theories with a large gap $\Delta_1$ to the lowest nonzero dimension of a primary field.  Clearly,
if there exist two such theories, $H$ must be at or above the gap.
Simple
arguments show that, for a chiral CFT, the gap cannot be larger than $\Delta_1=c/24+1$, and
this bound is saturated for $c=24$.
Although it is not known whether such ``extremal CFTs'' exist for $c>24$, from works such
as \cite{Gaberdiel:2008xb} it seems 
plausible that $\Delta_1$ grows linearly with $c$.  This suggests the growth of $H$ proposed
in the conjecture.

While some property like this seems necessary for a theory space as we are envisioning it to
exist at all, in itself it would not answer our previous questions.  For example, it would not
directly imply any bound on the number of such operators in a given CFT.  
Nevertheless we feel this would be an important statement about the space of CFTs.

\subsection{Convergence and relation to comparing partition functions}

To fix the problem we discussed in the previous subsection,
we can put in a convergence factor which suppresses the contributions
of high dimension operators.  Thus, instead of \eq{normdist}, we use
\bea \label{eq:rdist}
d(T_1,T_2) &=& \bigg( \sum_i |e^{-t 2\Delta_i^{(1)}} \Delta_i^{(1)}- e^{-t 2\Delta_i^{(2)}} \Delta_i^{(2)}|^p   \\
&+& \sum_{i,j,k} |e^{-t (\Delta_i^{(1)}+\Delta_j^{(1)}+\Delta_k^{(1)})} C_{ijk}^{(1)}
  - e^{-t (\Delta_i^{(2)}+\Delta_j^{(2)}+\Delta_k^{(2)})} C_{ijk}^{(2)}|^p 
\bigg)^{1/p} \nonumber
\eea
where $\Delta_i=h_i+\bh_i$, and
with some fixed $t>0$.  This particular choice of convergence factor is made to realize the
axioms \eq{symmetry} and \eq{triangle}, and
can be motivated in
terms of the world-sheet origin of the data $h_i$ and $C_{ijk}$.  For example, $C_{ijk}$
corresponds to an amplitude on a sphere with three punctures, and the regulated version
is obtained by replacing each puncture with a boundary
of length $2\pi t$ (obtaining the ``pair of pants'' diagram).  Defining the operator
\be
H \equiv \sum_a L_0^{(a)} + \bL_0^{(a)}
\ee
where the sum over $a$ represents a sum over the boundaries, we can write this as
\be \label{eq:cnormdist}
d(T_1,T_2) = \min_U ||U \; e^{-t H_1} V(T_1)- e^{-t H_2} V(T_2)|| ,
\ee
where we reintroduced the unitary transformation of \eq{Udistance}.

This type of definition could also be expressed in terms of differences between the
correlation functions of the two CFTs.  By expanding in a sum over intermediate states,
we can write a four-point function as
\be
\vev{\phi_1(x_1)\phi_2(x_2)\phi_3(x_3)\phi_4(x_4)} = f(x) \sum_i C_{12}^i C_{34i} e^{-t h_i-\bt \bh_i} ,
\ee
where $(t,\bt)$ are simple functions of the $x_a$ (essentially the logarithms of cross-ratios), as is $f(x)$.
At fixed $t$ one gets a regulated sum of o.p.e. coefficients of the form we want, and clearly one
could use the $t$ dependence to extract the coefficients and construct an expression like \eq{rdist}.

This still leaves the need to minimize over unitaries $U$.
One could reduce these issues by, instead of comparing individual correlation functions and
then adding the results, taking a physical observable which already combines all the
o.p.e. coefficients.  Consider the
genus two partition function; by thinking of this as obtained by attaching two ``pairs of pants''
one sees that it has an expansion of the form
\be
Z_{g=2}(T;\; t_1,t_2,t_3) = \sum_{ijk} |C_{ijk}|^2 e^{-t_1 \Delta_i-t_2 \Delta_j-t_3 \Delta_k}
\ee
and could also be used for this purpose.  We will return to this idea shortly.

\subsection{More physical definitions}
\label{ss:phys-def}

Some of these mathematical prescriptions can be translated into more physical terms.
A prototype for this is to consider boundary conditions in a single CFT.  Given a pair
of boundary conditions $B_1$ and $B_2$, a natural way to define the distance between
them is in terms of the lowest dimension open string operator between them
\cite{Douglas:1997ch,Douglas:1999ge,Moore:2003vf}.  
Given the partition function $Z_{12}$ for such open strings,
we define
\be \label{eq:d-geom}
d^2(B_1,B_2) = \frac{c}{24}-\lim_{t\rightarrow\infty} \log Z_{12}(t)
\ee
in conventions in which an operator of dimension $h$ contributes $\exp (c/24-h)t$ to $Z$.
The relation $d^2=h$ is motivated by the standard formula for the dimension of a winding
state in the large volume limit.

This will be zero if $B_1\cong B_2$, since the identity operator will contribute.
Naively it can also be zero if $B_1\ne B_2$
but the branes intersect, but in a unitary CFT the only operator of dimension zero
is the identity operator, which cannot exist in this Hilbert space.  
In fact there will be additional ``twist'' contributions to the ground state energy which
raise the dimension.  While this gives us two of the axioms, we do not know a proof of the 
triangle inequality for this definition (away from the large volume limit).

Let us now consider a distance between a pair of $d=2$ CFTs.  The direct analog of the
preceding discussion would involve a hypothetical $d=3$ theory which admitted $d=2$
CFTs as boundary conditions.  However another option, which
could be applied to any pair of CFTs $T_1$ and $T_2$,
would be to make some physical construction in the tensor product theory $T_1\otimes T_2$.  

We then need a physical analog of the unitary operator $U$ which relates the two
Hilbert spaces.  The simplest thing to consider would be the boundary states in the
product theory.  Equivalently, we can consider a defect line $L(U)$ which separates a region
of the world-sheet carrying theory $T_1$ from one carrying theory $T_2$ \cite{Bachas:2001vj}.
Given any world-sheet definition in which the two theories are related by perturbations and
flows, a precise definition of defect line can be made,
as discussed in many works such as \cite{Brunner:2007ur}.

If $T_1\cong T_2$, we can find a ``topological boundary state'' which can be freely moved
without changing the energy; in this sense it has zero tension.
This suggests that we define $d(T_1,T_2)$ in terms of the
tension of the defect, minimized over all such defects.  One must decide whether to use the
leading nonuniversal term in the tension, or the more universal $g$-function.  The first
leads to a distance of the general type \eq{cnormdist}, but using the
option we discussed earlier of always taking $||V||=1$.  The second will be discussed elsewhere
\cite{BD}.  By analogy to \eq{d-geom}, one can take $d^2(T_1,T_2)=\log g(T_1,T_2)$, and
realize most of the axioms, but the triangle inequality is not manifest.

Let us now return to \eq{cnormdist}.  For the special case $p=2$, we can write the
norm $||V||$ as the square root of an inner product $(V,V)$.  Now, making the world-sheet
identifications of the previous subsection, the inner product on the CFT Hilbert space $\CH$,
gives us an inner product on the space of coefficients $V$ (for example, $C_{ijk}$ lives in the
dual of $\CH\otimes \CH\otimes \CH$, etc.).  Thus, we can rewrite 
the regulated difference between the o.p.e. coefficients appearing in \eq{cnormdist} as
\bea \label{eq:wsdistance}
d_C(T_1,T_2)^2 &=& \min_U (U\; e^{-t H_1} V(T_1)- e^{-t H_2} V(T_2),U \;e^{-t H_1} V(T_1)- e^{-t H_2} V(T_2)) \\
 &=& ||e^{-t H_1} V(T_1)||^2 + ||e^{-t H_2} V(T_2)||^2 - 2 \max_U ( U\; e^{-t H_1} V(T_1), e^{-t H_2} V(T_2)) 
 \nonumber \\
 &=& Z_{g=2}(T_1\; 2t,2t,2t) + Z_{g=2}(T_2\; 2t,2t,2t) - 2 \max_U Z_{g=2}(T_1|L(U)|T_2;\; 2t,2t,2t) 
 \nonumber
\eea
where the last term 
\be \label{eq:Udefect}
Z_{g=2}(T_1|L(U)|T_2\; 2t,2t,2t) = (U \;  e^{-t H_1} V(T_1), e^{-t H_2} V(T_2))
\ee
is defined by these equations.  Now, the point is that this term also has a world-sheet 
interpretation.  Since the unitary $U$ maps $\CH_1$ to $\CH_2$, we
realize it as a defect line $L(U)$, separating two regions of the world-sheet carrying the two
theories.  Thus it is the partition function on a genus 2 Riemann surface, built by
sewing together two pairs of pants, but with the common defect line $L(U)$
along all three components of the junction.  

A similar discussion shows that the regulated difference between the operator
dimensions in the two theories can be written as the $d/dt$ derivative 
(or equivalently an insertion of $L_0+\bL_0$) of a
difference of genus one amplitudes,
\be \label{eq:wsonedistance}
d_h(T_1,T_2)^2 = -\frac{d}{dt} \left( Z_{g=1}(T_1; 2t) + Z_{g=1}(T_1; 2t) - 2 \max_U Z_{g=1}(T_1|L(U)|T_2;\; 2t) \right) .
\ee

Finally, the squared distance $d^2(T_1,T_2)$ is defined as the sum of
\eq{wsdistance} and \eq{wsonedistance},
\be \label{eq:defect-distance}
d(T_1,T_2)^2 = \min_U  c_h(t)  d_h(T_1,T_2)^2 + c_C(t)  d_C(T_1,T_2)^2 
\ee
with the choice of defect line $L(U)$ which minimizes the total distance (so the argument around
\eq{triangle-proof} will work)
and where $c_h(t)$ and $c_C(t)$ are normalization factors chosen to cancel the leading divergences
of the partition functions (for example, $c_h(t) \sim \exp ct/6$).
By the previous formal arguments, this will satisfy the triangle inequality as well as the other axioms.
Thus we have a physical interpretation of the distance \eq{cnormdist} for the particular case $p=2$.

There are some evident variations on this one could consider.  For example, one could add
higher genus terms, or even integrate over moduli space (in the critical $c$ or with some suitable weight)
to make this look more like a string theory observable.  There are also
a number of questions about it which we hope to clarify in subsequent work.
One is whether to require that the defect lines be local on the world-sheet.
If so, then for \eq{cnormdist} and
\eq{defect-distance} to be the same, we would need to show that the minimizing $U$ is local.  

With some refinements, such as specifying the specific metric on the surfaces,
the definition \eq{defect-distance} might make sense for general 2d QFTs as well.  This point
is also under study.

\subsection{Topology and boundedness}

One role for a definition 
of distance between CFTs or QFTs is as a technical or mathematical tool, used
to establish properties like the convergence of an approximate definition.  We 
will discuss this idea in \S \ref{s:weak}.  But there are also interesting physical questions about it.
While some of these questions have already been asked for the Zamolodchikov metric,
because the definition of \eq{defect-distance} is essentially linear, 
it should be much simpler to study.

While the two metrics are different, one can hope that \eq{defect-distance} (or some
variation of it)
is uniform with respect to the Zamolodchikov distance.  In part, this
is to say that both metrics define the same topology:
a sequence of CFTs which converges in the Zamolodchikov metric, also converges
in the metric \eq{defect-distance}.  Similarly, if a sequence converges in \eq{defect-distance}, and all pairs
of CFTs in the sequence have well-defined Zamodchikov distances, it will converge there as well.

This is true and follows from a
more specific relation to the Zamolodchikov metric, which can be found by computing derivatives of
\eq{defect-distance} with respect to the moduli.  One has \cite{in-prog}
\be \label{eq:metric-rel}
-\frac{1}{2}\;\frac{\partial}{\partial t^i_1} \frac{\partial}{\partial t^j_2} d^2(t_1,t_2)\bigg|_{t_1=t_2=t} =
g^{(Zam)}_{ij}(t) \cdot Z(t)
\ee
where $Z(t)$ is the sum of partition functions appearing in \eq{defect-distance}.

Next, one wants the idea of ``infinite distance limit'' to be the same.  More precisely,
an unbounded sequence $T_n$ in one metric, for which the 
distances $d(T,T_n)$ to some reference theory $T$ go to infinity for large $n$,
should be unbounded in the other.  Whether
or not this notion is the same is a bit less clear, because of the factor $Z(t)$ in
\eq{metric-rel}.  This tends to increase the distance to the large volume limit.  But, since
this was already at infinite distance in the Zamolodchikov metric, the claim could still hold.

The boundedness property allows us to restate a conjecture of Kontsevich and Soibelman
(\cite{Kontsevich}, as cited in \cite{Acharya:2006zw}).  
Loosely stated, it is that the only infinite distance limits in the space of CFTs
are the large volume limits.  More precisely, it is that a family of CFTs with
fixed $c$ and a lower bound on the nonzero operator dimensions $h_i$, is totally bounded
(or precompact), meaning that it can be covered by a finite number of finite radius balls.

Besides determining the infinite distance limits, this implies that the space
can be completed, meaning that the limit of any convergent sequence of CFTs can be regarded
as defining a point in a larger completion of the space.  There are many examples of such limit points 
obtained by shrinking volumes of cycles to zero, whose interpretation in string theory involves
nonperturbative effects such as new massless states, instanton corrections, or both.
Such limit points need not satisfy all of
the axioms of CFT, and it would be interesting to have a general theory of them.
See \cite{Kontsevich,Roggenkamp:2003qp,Roggenkamp:2008jm,Soibelman} for work along these lines.

As a final comment, although we arrived at a nice definition with a concrete 
world-sheet interpretation, in mathematics one often uses more than one definition
of distance and/or norm, even in the same problem, as each can have its own advantages with respect
to the other structures in the problem.

\section{Embedding theorems}

Any approach based on observables such as the above faces the problem that the consistency
conditions for observables to come from a QFT are very complicated.  On the other hand, the
consistency conditions on a Lagrangian which can be used to define a QFT by a functional
integral are fairly simple.  This suggests that we base the classification on the Lagrangian
approach, however we do not know whether all QFTs can be obtained this way.

A very direct way to address this question would be to give a prescription which, given some
subset of the observables of a QFT, produces a Lagrangian which is guaranteed to reproduce
the original QFT.  While at first this may sound implausible, what I have in mind is a prescription
which is highly redundant, and perhaps only useful to make this conceptual point.  For example,
we might imagine a prescription in which every operator up to some very high dimension is
represented by its own field in the Lagrangian.  Let us consider the Landau-Ginzburg theories
with a single real field $\phi$ and the Lagrangian
\be
\CL = (\partial\phi)^2 - \sum_{i=1}^n t_i \phi^i ,
\ee
then we would introduce operators $O_k=:\phi^k$ up to $k=n-2$, and possibly further
operators involving derivatives of $\phi$.  We would then introduce couplings which
enforce the relations between these fields, such as Lagrange multipliers
\be
\lambda (O_2 - C_{112} O_1^2)
\ee
where $C_{112}$ is an o.p.e. coefficient, and so on.
The idea would be to have so many couplings that we could tune as many
dimensions and o.p.e. coefficients as we needed to reproduce the desired QFT.

Is this a reasonable goal?  Even if it is possible, it might turn out to have the problem we
mentioned in the introduction, that even after putting some upper bound on the number
of degrees of freedom of our QFT, we might need an arbitrarily large number of fields
in order to reproduce it.  If so, this approach would probably not be worth the trouble to
develop.

Restricting attention to
2d CFT, we are thus led to conjecture the ``inverse $c$-theorem''
of the introduction, 
\begin{conjecturetwop} \label{q:two}
All 2d CFTs of central charge $c$ can be obtained
as flows from a theory of $N$ free bosons (possibly with gauge fields), for some finite $N$ 
depending on the central charge $c$.
\end{conjecturetwop}

In other words, every 2d CFT can be realized as a linear sigma model.
What reason is there to believe this, and what should we expect $N$ to be?  There are many
analogous questions in mathematics which can guide us.

Suppose we want to classify the
manifolds of a given dimension $D$.  Might it be that all such manifolds can be obtained
as submanifolds of $\BR^N$, the Euclidean $N$-dimensional space?  More explicitly, we
are asking for an embedding, a function $X:M\hookrightarrow\BR^N$ such that $X(p)=X(q)$
implies $p=q$, and the Jacobian $\partial_i X^I$ has maximal rank $D$ at each point.

This question was
asked and answered early on; for smooth manifolds this is the Whitney Embedding Theorem,
and one needs $N=2D$.  Thus, if all CFTs could be obtained as sigma models with target
spaces of dimension equal to the central charge, and we did not need to tune the metric
or couplings, the conjecture would hold with $N=2c$.

This is of course a drastic simplification of the real situation.  A closer analog might be
to ask that we can reproduce the relevant and marginal couplings of the sigma model,
meaning the potential, metric and $B$-field via the embedding.  Denoting these as $(V,g_{ij},B_{ij})$,
the mathematical
question is to find a function $f$ on some $\BR^N$ such that $f=0$ on
an embedding $X:M\hookrightarrow\BR^N$ and is positive elsewhere, so that
$$
g_{ij} = \partial_i X^I \partial_j X^J \delta_{IJ} ; \qquad
B_{ij} = \partial_i X^I \partial_j X^J \omega_{IJ}
$$
on $M$, where $\omega_{ij}$ is a constant nondegenerate antisymmetric tensor.
The total potential would then be $C f + \tilde V$, where $\tilde V$ is a
function on $\BR^N$ which restricts to $V$ on $M$, and $C$ is some large constant.

Looking into the mathematics, finding $f$ and $\tilde V$ is not hard.\footnote{
One can
take for $f$ the sums of squares of functions which define $M$ locally, and combine these
with $V$ using a partition of unity on $\BR^N$.}
To realize the metric as the restriction of the Euclidean metric
on $\BR^N$, and the $B$-field as the pullback of a constant two-form, since together these
have $D^2$ coefficients, one might think that it would be possible given $D^2$ adjustable functions,
so that $N=D^2$.

Showing that this can be done globally ({\it i.e.} over all of $M$) is much more difficult.
Nevertheless, the Nash embedding theorem states that this is possible for a smooth metric;
one needs $N = (D+2)(D+3)/2 $ (see \cite{Andrews} for a recent review).  While I don't know of a
similar theorem for the combined $g$ and $B$ problem, if $H=dB=0$, 
there is no obstruction to realizing $B$
as a pullback, so it seems plausible that this can also be done with $N$ of order $D^2$.

This suggests that the conjecture could hold with $N$ of order the square of the
central charge $c$.\footnote{
This is the conjecture as stated in the QTS6 proceedings.}
On the other hand, the restriction to $H=0$,
which follows because $d$ commutes with pullback, is significant.  
How can we get WZW models, or other models with $H\ne 0$, via embedding?

In fact, I know of no theorem stating that certain CFTs require $H$ field for their definition.
The basic counterexample is the $SU(2)$ WZW model at level $k=1$, which is equivalent
to a free boson at the self-dual radius.  Since one can realize larger $k$ by taking sums of
these models and gauging \cite{Gawedzki:1988nj}, Conjecture $2^\prime$ does not seem to be ruled out,
but we lose our guess for $N$.
In addition, $H$ field is an important aspect of CFT and a good embedding would
realize it in a more manifest way.

Another option is to generalize the conjecture and allow flows from ``almost free theories'' 
with $H\ne 0$.  In particular, the WZW models are exactly solvable, in a way that can be
understood geometrically in terms of parallelizing torsion.  Thus, granting that the
appropriate embedding theorem exists in the large volume limit, this suggests 
\begin{conjecture} \label{q:twoprime}
All 2d CFTs of central charge $c$ can be obtained
as flows from a gauge theory of free bosons coupled to $SU(2)$ WZW models, with a total
number of factors
growing as the square of the central charge $c$.
\end{conjecture}

Finally, a third option might be to take better advantage of the gauging, as cohomologically
nontrivial forms can be produced this way.  This is how the K\"ahler form is realized in the
$(2,2)$ gauged linear sigma models \cite{Witten:1993yc}.  However, while topologically nontrivial
$3$-forms can be realized this way, I don't know a way to do this starting with a free theory.

In any case, one needs to discuss renormalization and many other issues
to follow this geometric route.  Clearly the problem will be far more constrained given supersymmetry.
For the case of $(2,2)$ SCFT, the relevant analogy is to the Kodaira embedding theorem, 
which suggests a general prescription for taking data from the topological open sector of an SCFT, and
constructing a gauged linear sigma model which flows back to the same SCFT \cite{AD}.

\section{Weak QFTs} \label{s:weak}

Our final approach is to try to
sit the problem of constructing and classifying QFTs, in some larger problem obtained
by relaxing some of the many axioms and consistency conditions.  The hope is of course
that this problem will be easier to solve, and that we can then go on to enforce the
additional conditions required to get actual or ``honest'' QFTs.  
In other words, we start by classifying ``weak QFTs'' which obey
some but not all of the axioms of QFT.  

The term\footnote{suggested to me by Edward Witten}
is by analogy to the idea in mathematics of a ``weak
solution'' of a partial differential equation.  This is a solution which need not be a function
but might be a distribution or something else.  Even if the goal is to find a solution which is
a function, often it can be easier to first show that such a weak solution exists, and then show
that it is continuous, smooth or satisfies other desired properties.  The basic example is
the solution of linear PDE's by Fourier transform, which does not naturally produce a function
but rather an ``$L^2$ function,'' which is only defined up to a set of measure zero.

When stated in such generality, of course this approach has been used extensively, one might
even say universally, in the problem of defining QFTs.  For example, a lattice theory or a 
regulated functional integral might be regarded as defining a ``weak'' form of a Euclidean
or Lorentz invariant local QFT.  Of course, in this case one has good reasons to expect the desired
symmetry and locality properties to be recovered in the long distance limit.  

The class of weak theories we have in
mind here would be one in which we can define the weak QFT in terms of a finite amount of data,
and then express the additional axioms needed for it to be an honest QFT as some finite set
of equations which the data must satisfy.

A prototype for this is the classification of the minimal $c<1$ CFTs.  Such a CFT is defined
by giving a finite list of primary fields and the finite set of their o.p.e. coefficients.  This data then
must satisfy the BPZ (associativity) and Cardy conditions to define an honest CFT.

It would be interesting to be able to generalize this to $c>1$.  The main problem is of course that
we cannot work with the actual list of primary fields.  We need to know something like 
Conjecture \ref{q:one} to cut this down to size.

Suppose we knew the data of Conjecture \ref{q:one}; {\it e.g.} the complete list of primary
fields and o.p.e. coefficients for all operators with dimension $h<H$ and $\bh<H$.  Could
we use this to define a ``weak QFT'' ?

An interesting related idea appears in the work \cite{Rattazzi:2008pe}
of Rattazzi {\it et al}.
They show that a CFT with a scalar operator $\phi$ of dimension $h_\phi$, must contain
an operator $:\phi^2:$ in the o.p.e.
\be \label{eq:phisqope}
\phi\, \phi \rightarrow 1 + :\phi^2: + \ldots
\ee
with an upper bound on its dimension $h_{\phi^2}\le f(h_\phi)$,
where  $f(h_\phi)\gtrsim 4h_\phi$ in $d=2$.  Interestingly,
the argument does not use Virasoro representation theory, only the finite-dimensional conformal
group $SO(d,2)$, and thus
could work in any space-time dimension.  One needs explicit expressions
for the conformal blocks,  which were found for $d=2$ and $d=4$ in
\cite{Dolan:2003hv}.
 
The argument relies on associativity of the 4-point function, schematically
\be \label{eq:assoc}
\sum_i \vev{\phi(z_1)\phi(z_2)|i}\, \vev{i|\phi(z_3)\phi(z_4)} = 
\sum_j \vev{\phi(z_2)\phi(z_3)|j}\, \vev{j|\phi(z_4)\phi(z_1)} .
\ee
It is important that every term on both sides is positive, by unitarity (reflection positivity).
Using conformal symmetry, one can write this (in any $d$) as a relation between functions of one
complex variable $z$, a generalization of the cross ratio in $d=2$ which maps the coincident limits
$z_1\rightarrow z_2, z_3, z_4$ to $z\rightarrow 0,\infty,1$ respectively.  Then,
while this is a functional relation, one focuses on the behavior at the symmetric point
$z=1/2$.
Although it is not at all obvious, from numerical study
it turns out that the difference of the conformal
blocks entering the two sides of this equation, behaves differently at $z=1/2$ for
intermediate operators $O_i$ with $h_i < f(h_\phi)$ and those with $h_i > f(h_\phi)$.
This can be expressed as a sign of a second derivative which must vanish for \eq{assoc}
to hold, and thus both types of operators must be present in the sum.\footnote{
See \cite{Hellerman:2009bu} for a similar argument at genus one.}

In the terms we are using here, one can think of this as a ``weak CFT'' argument.
One is considering a class of theories which has most of the structure of CFT, such as
symmetry and positivity, but need not satisfy associativity.  Within this class, one can
write down four-point functions which match an assumed part of the spectrum and o.p.e.
\eq{phisqope}.  One then imposes associativity and gets a constraint.

Let us return to our earlier question:
how can we use the data of Conjecture \ref{q:one} to define a ``weak QFT'' ?
What we clearly can do, is construct approximate four-point functions by summing
over intermediate operators with $h,\bh\le H$, such as
\be
\sum_{m\, \mbox{such that}\, h\le H} \vev{\phi_i(z_1)\phi_j(z_2)|m}\, \vev{m|\phi_k(z_3)\phi_l(z_4)} 
\ee
While these will reproduce the data, they will not be associative.  

Then, we define a function $U$ which expresses how far these
approximate four-point functions are from associativity, schematically
\be
U = \sum ||F - F'|| 
\ee
where $F-F'$ is the difference between the two sides of \eq{assoc}, and the sum is over all 
correlation functions of operators with $h,\bh\le H$.

Of course, associativity of the four-point functions is not the only consistency condition
on a CFT, there is also modular invariance at genus one,
and perhaps others (since we did not impose
associativity of all four-point functions).  But we could go on and add
other terms to measure the failure of these properties to hold.

We can then define a flow which decreases $U$.  Given a choice of parameters to vary, say
the dimensions and o.p.e. coefficients, and given a metric on this space, we have the
gradient flow
\be
\dot V^i = - G^{ij} \frac{\partial U}{\partial V^j} .
\ee
Thus, starting with a weak QFT, we flow to an honest QFT.

This scheme is admittedly somewhat complicated.  Worse,
it has two major problems:
\begin{itemize}
\item If we do not put in all the consistency conditions, we may end up at $U=0$ with
a set of correlation functions which do not correspond to any QFT.
\item The function $U$ might have local minima with $U>0$.  Indeed,
as we put in more and more consistency conditions, the function
$U$ becomes complicated, and is likely to have more and more local minima.
\end{itemize}

So, we should probably not rush to implement this scheme numerically just yet.  But the
general idea might be a good one, analogous to constructing CFTs via RG flow but
making different aspects of the problem manifest.
Given enough analytic understanding of the space of CFTs one is trying to describe, one
might be able to make good choices for $U$ and the metric which avoid the problems we
just raised.

\bigskip \noindent
{\it Acknowledgements}
\smallskip

Of the many people I have discussed this topic with over the years,
I would particularly like to thank Maxim Kontsevich and Edward Witten, who introduced me to many
of these ideas.  I also thank Costas Bachas for discussions on interfaces and the 
ideas of \S \ref{ss:phys-def}.

This research was supported in part by DOE grant DE-FG02-92ER40697.

\end{document}